\documentclass[10pt,a4paper,twocolumn]{narms} 

\usepackage{subfigure}
\usepackage{psfig,epsfig}
\usepackage{timesmt}
\usepackage{chicaco} 
\usepackage{amssymb,amsmath}

\makeatletter
\renewcommand{\section}{\@startsection
{section}
{1}
{0mm}
{- \baselineskip}
{0.15\baselineskip}
{\normalfont\normalsize}}

\renewcommand{\subsection}{\@startsection
{subsection}
{2}
{0mm}
{-\baselineskip}
{0.15\baselineskip}
{\normalfont\normalsize}}
\makeatother

\begin{document}

\title{Free volume distributions inside a bidimensional granular medium}
\author{\large {F. da Cruz$^1$, F. Lechenault $^{1 *}$, O. Dauchot$^1$ and E. Bertin$^2$}
\\
{\em $^1$Commissariat a l'Energie Atomique, F-91191 Saclay, France}
\\
{\em $^2$Department of Theoretical Physics, University of Geneva, CH-1211 Geneva 4, Switzerland}
}
\date{}

\abstract{We investigate experimentally within a two-dimensional cell the distribution of the free volume associated either to single grains or to clusters of grains. Our main result is that the logarithm of the probability to find a free volume per grain $v_N^f$ in a cluster of $N$ grains scales as
$N^{\alpha} g(v_N^f)$, with $\alpha = 0.75$. We interpret this non extensive scaling factor $N^{\alpha}$ as an evidence for the onset of long range correlations between the free volumes of individual
grains. We also discuss the possible relation between $g(v_N^f)$ and Edwards entropy.
}

\maketitle
\frenchspacing

\section{INTRODUCTION}

Everyday life and recent experiments suggest that a thermodynamical description of granular media might be feasible~\cite{Nowak98,Schroter05}. Given that granular media consist in a large number of grains, there is a strong motivation for providing a statistical ground to this hypothetic thermodynamical description. However, the stationary dynamics of granular media results from the balance of dissipation and forcing by a non thermal source, so that  it does not leave any known ensemble invariant and little is known on the typical configurations explored dynamically~\cite{Barrat00}.

Still, it has been argued by Edwards and collaborators~\cite{Edwards89,Mehta89} that the dynamics is controled by the mechanically stable --the so-called blocked-- configurations and that all such configurations of a given volume are statistically equivalent provided that the driving involves extensive manipulations, such as shaking, shearing or pouring. This immediately leads to the definition of a configurational entropy $S_{conf}(V)$, and the associated state variable, the "compactivity" $X_{conf}^{-1}=\partial S_{conf}/\partial V$.

First attempts to test this flat measure assumption have taken advantage of the analogy between the gentle compaction of grains and the aging of glassy systems~\cite{Liu98,DAnna01}. Despite some clear examples where Edwards' approach fails~\cite{Godreche05}, explicit checks have been made so far in mean field models of the glass transition~\cite{Monasson95}, in schematic finite-dimensional models with kinetic constraints~\cite{Barrat00}, in spin glass models with non-thermal driving between the blocked states~\cite{Dean01} and finally in a few more realistic models of particle deposition~\cite{Brey00} or MD simulations of shear driven granular media~\cite{Makse02}.

However, clear evidence in real granular media is still lacking. A steady state, history independent dynamics under vibration has been observed by Nowak et al.~\cite{Nowak98}, for tapped glass beads at rather high volume fraction. More recently, a one to one correspondance between the volume fraction fluctuations and the volume fraction itself have been observed, so that assuming the validity of the usual thermodynamical relation $<\delta V^2>=X^2 \partial <V>/\partial X$, a measure of the compactivity $X$ as a function of the volume fraction has been derived~\cite{Schroter05}.

It would now be of major interest to produce further experimental evidence of the statistical foundation of such thermodynamical properties. In the present paper, we report experimental results on the free volume distributions inside a bidimensional granular  medium. We show that the probability distribution of the free volume per grain inside a cluster of N grains follows:
$$P(v^f_N)=\frac{1}{Z}e^{-N^\alpha \left(\frac{v^f_N}{X}-s(v^f_N)\right)},\quad {\rm with}\quad \frac{1}{X}=\frac{\partial s}{\partial v} \Big\vert_{\langle v \rangle}.$$

\section{EXPERIMENTAL SET-UP }
\label{experiment}
\begin{figure}[!ht]
 \begin{minipage}{\linewidth}
   \epsfig{file=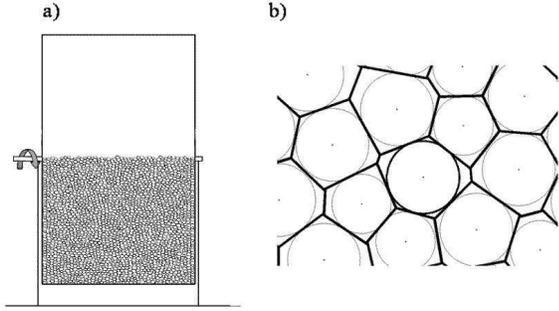,width=\linewidth}
   \caption[]{Experimental set-up and sketch of the modified Voronoi tessellation. }
  \label{flipflap}
 \end{minipage}
\end{figure}

The experimental cell (Fig.~\ref{flipflap}a) consists in a $60 {\rm cm}$ by $50 {\rm cm}$ rectangular glass container with an inside gap of thickness $3.5 {\rm mm}$. It contains 5000 nickel plated brass cylindrical spacers of two different sizes in equal number. Both kind of spacer is $3$ mm thick, and their diameters are respectively $d_s = 4 {\rm mm}$ and $d_l = 5 {\rm mm}$. In the following $d_s$ has been chosen as the unit length, so that the respective area of the small and large disks are $v_s^0=\pi/4\simeq 0.78$ and $v_l^0=\pi/4(d_l/d_s)^{2}\simeq 1.23$. The cell is half filled with a single layer of such hard disks mixed together, resulting in an homogenous and disordered bidimensional packing. The cell is mounted on a horizontal axis and rotated around this axis in such a way that the grains fall from one side to the other every half cycle.

The experimental procedure is the following. The cell starts in a vertical position and is rotated of one cycle, at a constant speed of one cycle per minute. During this cycle, the grains fall from one side to the other and then back to the initial side. We then stop the engine, wait for a few seconds to allow the system to reach a mechanically stable state, and take a picture of the bulk with a CCD camera. We perform 15 000 of such cycles.

The pictures hence taken display on average 300 grains. For each picture the centers of the spacers are located and their Voronoi diagram is computed (Fig~\ref{flipflap}b), taking into account the bidispersity of the assembly. We then collect the area of the cells along with the type, position and index of the associated grains. Out of these raw data, we extract and analyze the statistical distribution of the volumes occupied first by one grain, then by clusters of an increasing number of neighboring grains.

\section{DATA ANALYSIS }
\subsection{ {\em One grain volume distribution}}

\begin{figure}[!ht]
 \begin{minipage}{\linewidth}
   \center   
   \epsfig{file=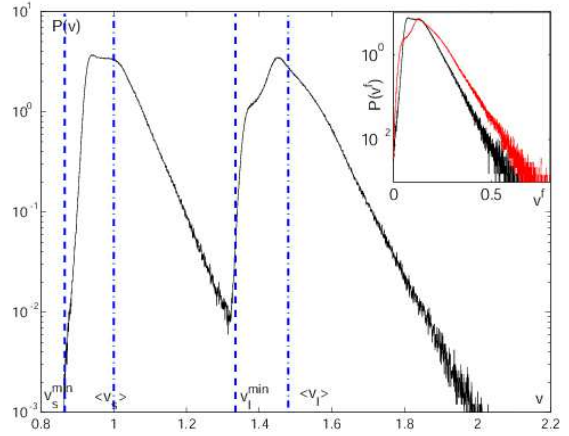,width=0.9\linewidth}
   \caption[]{Distribution of the Voronoi cells area. Vertical dashed lines : minimal Voronoi cell area. Vertical dash-dotted lines : conditional average Voronoi cell area. Inset: distributions of the free volume conditioned by the grain size; (dark): small grains; (grey): large grains}
  \label{pdf_v_vf}
 \end{minipage}
\end{figure}

Figure~\ref{pdf_v_vf} displays the distribution of the Voronoi cell area. The distribution displays two peaks centered on $<v_s> = 1.00$ (resp. $<v_l> = 1.49$), the averaged area occupied by the small, (resp. the large) grains computed independently. Also indicated on the figure are the minimal values that a Voronoi cell can possibly take - the closest regular hexagon- for each type of grain, $v_s^{min}=\sqrt{3}/2 \simeq 0.866$ and $v_l^{min}=\sqrt{3}/2(d_l/d_s)^2\simeq 1.35$. Both peaks present a well defined exponential tail, which is easily isolated when considering the distributions of the free volume ($v_{s,l}^f=v-v_{s,l}^{min}$), for each type of grain as shown on the inset of figure~\ref{pdf_v_vf}. It shall be noticed that the characteristic free volume for each type of grain, ${v_s^f}^*= 0.055$ and ${v_l^f}^* = 0.060$ are closer to each other than one would have expected given the particule size ratio. In other words, the characteristic free volume accessible to one grain not only depends on the size of that grain, but also on the neighborhood, most often composed of grains of both sizes.
\begin{figure}[!ht]
 \begin{minipage}{\linewidth}
   \center
   \epsfig{file=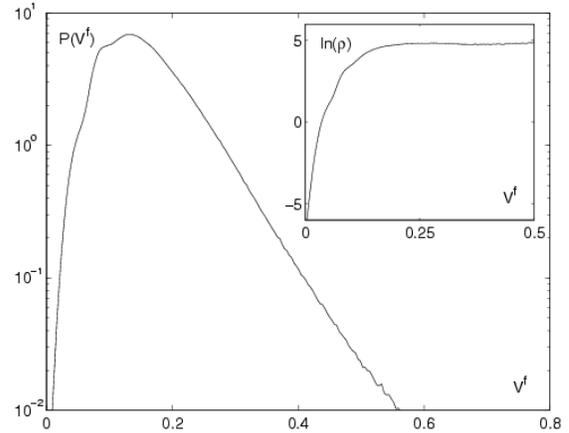,width=0.9\linewidth}
   \caption[]{Distribution of the free volume including both kinds of grains. Inset: density of states of one grain surrounded by a free volume $v^f$ (see text for details). }
  \label{pdf_vf1}
 \end{minipage}
\end{figure}

Figure~\ref{pdf_vf1} displays the distribution of the free volume without conditioning it by the type of grains.  One still observes the well defined exponential tail, with a characteristic free volume ${v^f}^*=0.058$. In the spirit of Edwards' description, one may assimilate this volume, characteristic of the sample, to a compactivity. This is of course a rather strong interpretation of the data, since it would be necessary at least to check that the dependence of the distributions on this compactivity is indeed fully embedded in the exponential tail. Still, as an indication, we have plotted in the inset of figure~\ref{pdf_vf1} the quantity $\ln(\rho(v^f))=\ln(P(v^f))+v^f/{v^f}^*$, the density of states of one grain surrounded by a free volume $v^f$. It saturates for large free volumes, after a sharp increase for small values.

At this point, it should be noticed that in granular media, one may not expect the N-body probability distribution for the volumes of a set of N grains to factorize as a product of one-grain distributions, because of possible correlations and degeneracies of the free volume states. In order to check if one recovers the neat properties of the usual canonical description for a large enough subsystem of the grain assembly, we hence turn to the volume distributions for clusters of neighboring grains. 

\subsection{ {\em Free volume distribution for clusters of neighboring grains}}

\begin{figure}[!ht]
 \begin{minipage}{\linewidth}
   \center
   \epsfig{file=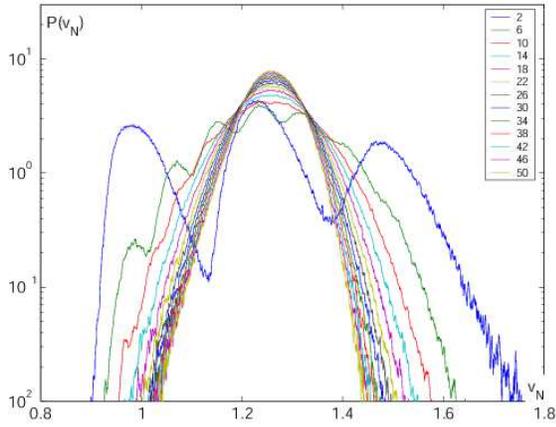,width=0.9\linewidth}
   \caption[]{Distribution of the volume per grain inside clusters of N grains: the larger N, the narrower the distribution}
  \label{pdf_vN}
 \end{minipage}
\end{figure}

\begin{figure}[!ht]
 \begin{minipage}{\linewidth}
   \center
   \epsfig{file=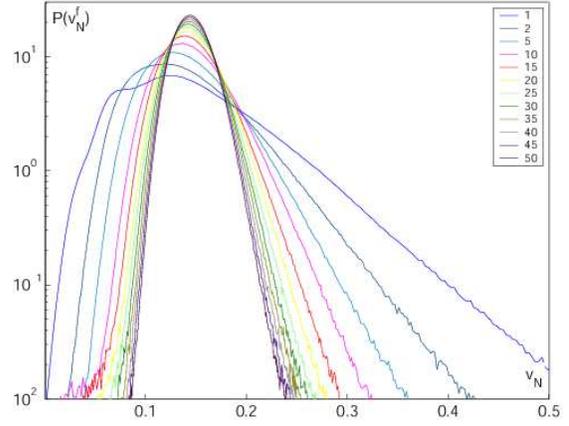,width=0.9\linewidth}
   \caption[]{Distribution of the free volume per grain inside clusters of N grains: the larger N, the narrower the distribution}
  \label{pdf_vfN}
 \end{minipage}
\end{figure}

Now dealing with clusters of grains, the free volume and the full volume are no longer equivalent, since clusters are composed of both kinds of grain. Figure~\ref{pdf_vN} (respectively figure~\ref{pdf_vfN}) displays the distribution of the volume (resp. the free volume) per grain inside clusters of $N$ neighboring grains, $v_N=N^{-1}\Sigma_{i=1}^N v_i$ (resp. $v^f_N=N^{-1}\Sigma_{i=1}^N v^f_i$).

The distribution of the free volume per grain $v^f_N$ converges much faster towards a well defined distribution than that of the volume. Indeed, it is clear from figure~\ref{pdf_vN}, in the case of the volume distributions, as $N$ increases from 1 to 10, that combinatory effects induced by the bidispersity give rise to series of peaks which complexify the distribution. For $N=2$, one observes three peaks, which are related to the three two-particles cluster configurations. Similarly, for $N=3$ --not shown here--, the distribution exhibits four peaks, etc. Such effects are absent in the case of the free volume distributions (figure~\ref{pdf_vfN}). Also, describing the volume distribution imposes to know the minimal value of the volume per grain, a non trivial quantity for clusters composed of grains of both sizes, whereas by construction, the distribution of the free volume readily addresses this issue. Accordingly, we restrict the following analysis to the free volume distributions.

We propose to describe the distributions of the free volume per grain inside a cluster of $N$ neighboring grains by a Gamma law of parameters $\eta_N$ and $X_N$ : 
$$P(v^f_N)=\frac{1}{X_N^{\eta_N}\Gamma(\eta_N)}(v^f_N)^{\eta_N-1}e^{-v^f_N/X_N}.$$ 
\noindent
where $\Gamma$ is the Euler Gamma function. This choice is motivated by the following considerations. 
In the hypothetic case where the one-grain free volume distribution would follow a Gamma law itself of parameter $(\eta_1, X_1)$, if the free volumes inside a cluster were independent variables, one would have exactly the proposed Gamma law for the distributions of $v^f_N$, together with the relations $(\eta_N=N \eta_1, X_N=X_1/N)$. This is a standard property, which seemingly extends for large enough $N$ in the case where the one-grain distribution is not a Gamma law~\cite{Bertin04}. Given that in the present case, the general shape of the one-grain free volume distribution already shares some of the key property of the gamma-law --being defined for $v^f>0$ and exhibiting an exponential dependence for large $v^f$--; given that the distributions seemingly belong to the same functional family irrespectively of $N$, it is reasonable to believe that the convergence towards the gaussian law expected in the limit of very large $N$ will occur via Gamma laws.

\begin{figure}[!ht]
 \begin{minipage}{0.50\linewidth}
   \epsfig{file=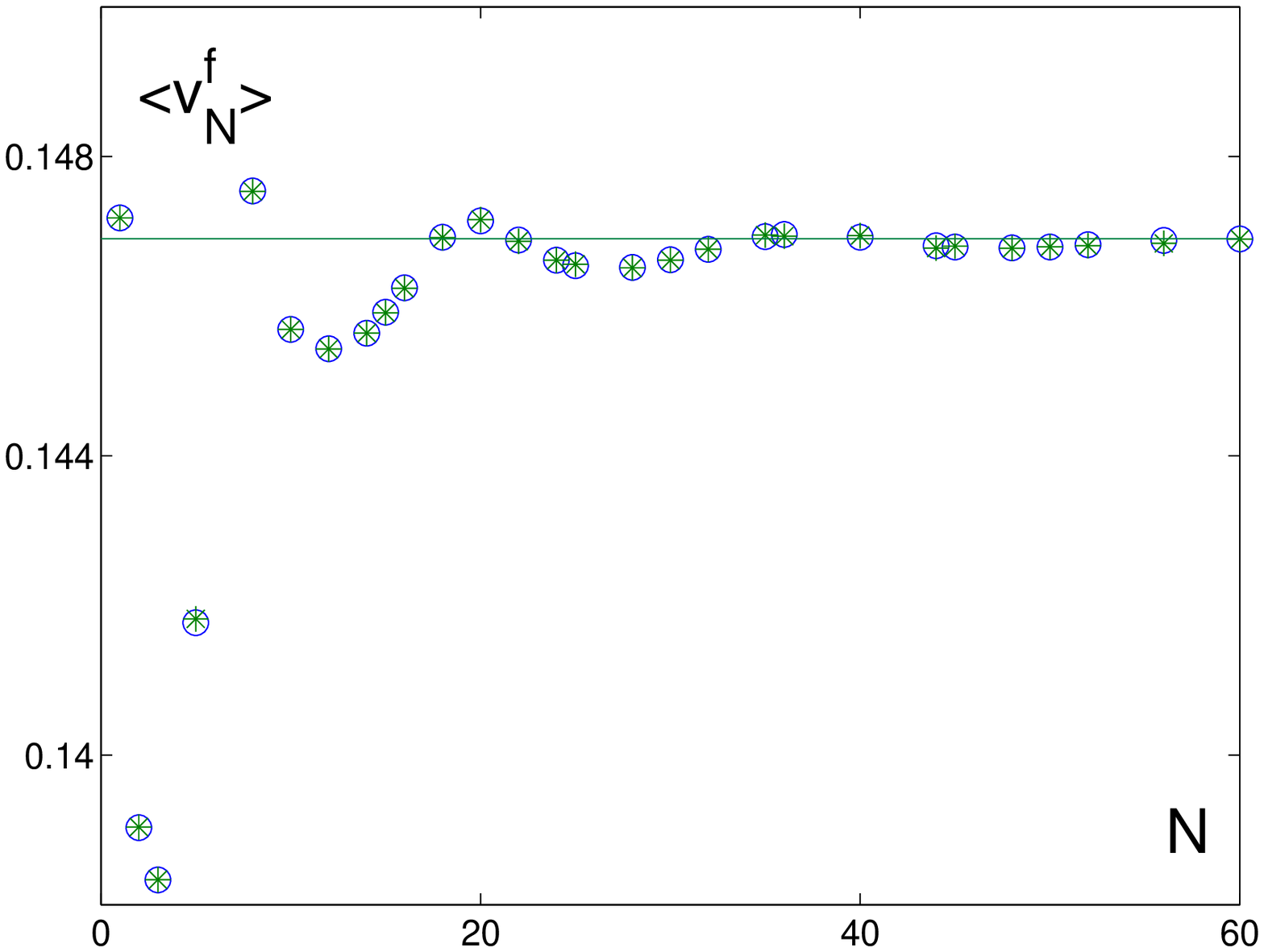,width=\linewidth}
 \end{minipage}
 \begin{minipage}{0.47\linewidth}
   \epsfig{file=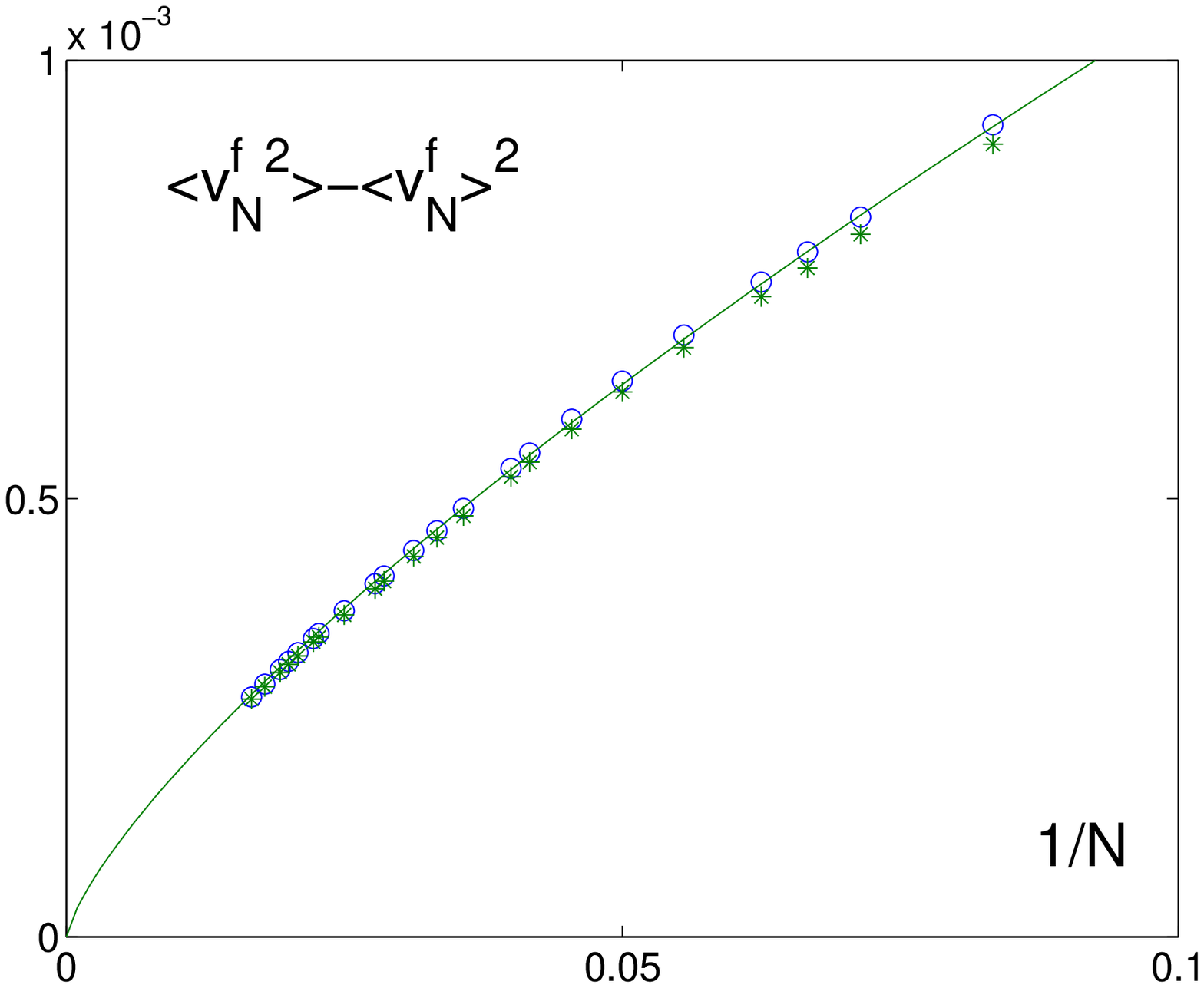,width=\linewidth}
 \end{minipage}
   \caption[]{Dependence on $N$ of the first (left) and second (right) moments of the free volume distributions ($\circ $): computed from the data; ($\ast $): extracted from the fit of the distributions by a Gamma law; (plain line): fit of their dependence on $N$}
  \label{moments_N}
\end{figure}

Once chosen the form of the distributions, we compute their first two moments and obtain $\eta_N$ and $X_N$, through the relations $<v^f_N>=\eta_N X_N$ and $<{v^f_N}^2>-<v^f_N>^2=\eta_N {X_N}^2$. As expected, $<v^f_N>$ rapidly evolves towards a constant (figure~\ref{moments_N}-left). On the contrary $<{v^f_N}^2>-<v^f_N>^2$ varies like $N^{-\alpha}$ with $\alpha=0.75\pm 0.0025$, in contrast with the $1/N$ dependence expected for independent variables (figure~\ref{moments_N}-right). 

As a direct consequence of the above analysis, the distribution of the free volume per grain inside clusters of $N$ grains is well described by a Gamma law, the parameters of which exhibit the following dependences on $N:$ $\eta_N=\eta_{eff} N^\alpha$ and $X_N=X_{eff} N^{-\alpha}$, with $\eta=3.5$ and $X=0.041$.  We have also plotted on these figures the dependences extracted from direct fit of the distributions by the proposed gamma-law; the results are the same. 

\section{DISCUSSION AND CONCLUSION}

Altogether, rewriting the above Gamma law in the limit of large $N$, one obtains that the logarithm of the distribution of the free volume per grain inside clusters of $N$ grains scales as $N^\alpha g(v,\eta_{eff},X_{eff})$ with $g(v)=\eta (\ln (v/(\eta X)) - v/(\eta X) + 1)$, $\alpha\simeq 3/4$, $\eta_{eff}\simeq 7/2$ and $X_{eff}=0.041$. This central result deserve a few comments.

In the case of Poisson Voronoi diagrams in two dimensions --where the centers of the Voronoi cells are randomly chosen--, it has been shown very recently~\cite{Jarai04} that the distribution of the Voronoi cell areas normalized by the mean area is precisely well described by a Gamma law of parameter $\eta=7/2$ and $X=1/\eta$. In the absence of correlations one obtains in the limit of large $N$, that the logarithm of the distribution of the volume per cell inside clusters of $N$ grains scales as $N g(v,\eta,X)$. Let us now compare our experimental results to this academic situations of ponctual grains with no hard sphere exclusion, and no correlations.

First, the observed non extensive factor $N^\alpha$ is presumably the evidence of long range correlations between the free volumes of individual grains. Indeed, in the presence of correlations decaying with the distance $r$ as $1/r^\gamma$, one has in two dimensions, for $\gamma<2$, that the second moment of the average of $N$ centered random variable scales like $N^{-\gamma/2}$. In the present case, we would thus infer the existence of long range correlations decaying like $1/r^{3/2}$.

Second the above analysis has allowed us to define effective parameters for the probability distribution of the free volume for one grain. Rather surprisingly $\eta_{eff}=\eta$: it seems that the hard sphere exclusion for the grains does not constraint the value of this parameter.

Finally, one can also write the distribution of the free volume per grain inside clusters of $N$ grains as:
$$P(v)=\frac{1}{Z}e^{-N^\alpha \left(\frac{v}{X}-s(v)\right)},\quad {\rm with}\quad s(v)=\eta (\ln\frac{v}{\eta X}+1),$$
and thereby $\frac{1}{X}=\frac{\partial s}{\partial v} \Big\vert_{\langle v \rangle}$, an exact result given the Gamma law distribution and more generally expected from a saddle point calculation in the large N 
limit.

Future work will have to check the robustness of the above results, when varying the type of grains or the dynamical protocole. Also, it will be of major interest to be able to vary the averaged volume fraction of the sample, in order to investigate the equation of state in the light of the above statistical description.

\bibliographystyle{chicnarm} 
\bibliography{bibGlass}
                   
\end{document}